\documentclass{aa}
\usepackage{amsmath}
\usepackage{graphicx}
\usepackage{pgfplots}
\usepackage{wasysym}
\usepackage{natbib}

\def\spin#1#2{%
\draw [domain=-120:120, thick,->] plot ({#1+0.2*cos(\x)}, {#2+0.2*sin(\x)});
\fill[] (#1, #2) circle (0.03);
}
\def\pole#1#2{%
\draw[thick] (#1, #2+-1.3) -- (#1, #2+1.3);
\draw [domain =-150:150, thick,->] plot ({#1+0.2*cos(\x)}, {1.33+#2+0.1*sin(\x)});
}
\def\toprockA#1#2{%
\fill[blue] (#1-1, #2) circle (0.15);
\node[color=blue] at (#1-1-0.3,#2) {\sf A};
}
\def\toprockB#1#2{%
\fill[color=red] (#1-0.3, #2) circle (0.15);
\node[color=red] at (#1-0.6, #2) { \sf B};
}
\def\botrockA#1#2{%
\fill[blue] (#1+1, #2) circle (0.15);
\node[color=blue] at (#1+1+0.3,#2) {\sf A};
}
\def\botrockB#1#2{%
\fill[red] (#1+0.25, #2+0.95) circle (0.15);
\node[color=red] at (#1+0.55,#2+1.1) {\sf B};
}
\def\seasons#1{%
    \node[] at (#1-2.5, -2) { spring };
    \node[] at (#1+2.5, -2) { summer };
    \node[] at (#1+2.5,  2) { autumn };
    \node[] at (#1-2.5,  2) { winter };
}

\def\imgcoors{%
\begin{tikzpicture}[scale=1.4]
    \fill[lightgray] (0,0) circle (1); 
    \draw[dashed] (0,0) ellipse (3 and 1.255);
    \draw[] (0,0) -- (2.8,1);
    \draw [domain=-48:20] plot ({0.5*cos(\x)}, {0.5*sin(\x)});
    \node[] at (0.3, -0.1) { $\upsilon$};
    \draw[thick,rotate=20] (0,0) ellipse (3 and 1.5);
    \draw[->,dotted] (0,0) -- (0,2);
    \draw[->,dotted] (0,0) -- (1.07,-1.175);
    \draw[->,dotted] (0,0) -- (2.92,0.3);
    \draw[] (1.05,-1.175) -- (3.0,-1.175);
    \draw [domain=0:31] plot ({1.105+0.6*cos(\x)}, {-1.175+0.6*sin(\x)});
    \node[] at (1.5,-1.07) { $\epsilon$ };
    \node[] at (1.1,-1.3) { $x$ };
    \node[] at (0, 2.15) { $z=z'$ };
    \node[] at (3.07, 0.3) {$y$};
    \fill[orange] (2.8,1) circle (0.2);
    \node[] at (3.05, 1.35) { Sun };
    \draw[->] (0,0) -- (-0.3*1.6, 0.6*1.6);
    \draw [domain=117:227] plot ({0.3*cos(\x)}, {0.3*sin(\x)});
    \node[] at (-0.15, 0.03) { $\psi$ };
    \draw[->,dotted] (0,0) -- (-1.1, -1.18);
    \draw[->,dotted] (0,0) -- (2.75, -0.5);
    \node[] at (2.9, -0.6) { $y'$ };
    \node[] at (-1.1, -1.3) { $x'$ };
    \draw [domain=227:313] plot ({0.4*cos(\x)}, {0.4*sin(\x)});
    \node[] at (0.0, -0.5) { $2\pi- \phi$ };
    \draw[->] (-0.3, 0.6) -- (0.1, 0.5);
    \draw[->] (-0.3, 0.6) -- (-0.05, 0.9);
    \node[] at (0.25, 0.5) {$y''$ };
    \node[] at (0.13, 1.0) {$z''$ };
    \node[] at (-0.5, 1.05) {$x''$ };
    \fill[brown](-0.3,0.6) circle (0.05);
\end{tikzpicture}
}

\usepackage{txfonts}
\def\mdp{\langle\hskip-2pt\langle\Pi\rangle\hskip-2pt\rangle}
\def\d{{\rm d}}

\begin{document} 
   \title{Obliquity dependence of the tangential YORP}

   \author{P. \v{S}eve\v{c}ek\inst{1}
          \and O.~Golubov\inst{2,3,4}
          \and D. J. Scheeres\inst{2}
          \and Yu. N. Krugly\inst{4}
          }

   \institute{Institute of Astronomy, Charles University, Prague, V Hole\v{s}ovi\v{c}k\'{a}ch 2, 18000 Prague 8, Czech Republic\\
              \email{sevecek@sirrah.troja.mff.cuni.cz}
         \and
             Department of Aerospace Engineering Sciences, University of Colorado at Boulder, 429 UCB, Boulder, CO, 80309, USA
         \and
             Kharkiv National University, 4 Svobody Sq., Kharkiv, 61022, Ukraine
         \and
             Institute of Astronomy, Kharkiv National University, 35 Sumska Str., Kharkiv, 61022, Ukraine
             }

   \date{Received 19 April 2016 / Accepted 20 May 2016}

% \abstract{}{}{}{}{} 
% 5 {} token are mandatory
 
  \abstract
  % context heading (optional)
  % {} leave it empty if necessary  
   {Tangential YORP is a thermophysical effect that can alter the rotation rate of asteroids and is distinct from the ``normal'' YORP effect, but to date has only been studied for asteroids with zero obliquity. }
  % aims heading (mandatory)
   {The tangential YORP force produced by spherical boulders on the surface of an asteroid with an arbitrary obliquity is studied.}
  % methods heading (mandatory)
   {A finite element method is used to simulate heat conductivity inside a boulder, to find the recoil force experienced by it.
   %Then the torque created by the boulders is numerically integrated over the surface of an ellipsoidal asteroid.
    Then an ellipsoidal asteroid uniformly covered by such boulders is considered and the torque is numerically integrated over its surface.
   %Ray tracing of incident and emitted rays is performed.
   }
  % results heading (mandatory)
   {Tangential YORP is found to operate on non-zero obliquities and decreases by a factor of 2 for increasing obliquity.}
  % conclusions heading (optional), leave it empty if necessary 
   {}


   \maketitle
%
%________________________________________________________________

\section{Introduction}
The Yarkovsky-O'Keefe-Radzievskii-Paddack (YORP) effect is the torque experienced by an asteroid
due to its asymmetric emission of incident light pressure radiation by its surface \citep{rubincam00,bottke06,vokrouhlicky15}.
It is useful to divide the YORP torque into two components.
Normal YORP (NYORP) is created by light pressure forces normal to the large-scale surface of the asteroid,
which do not cancel each other's torques because of large-scale asymmetries in the asteroid's shape \citep{rubincam00}.
Tangential YORP (TYORP) is created by light pressure forces tangential to the large-scale surface,
which are non-zero because stones lying on the surface emit different amounts of infrared light eastward and westward \citep{golubov12}.
Any asteroid is, to some extent, subject to both NYORP and TYORP, and the two effects are hard to disentangle from observations.

Of these two effects, TYORP has not been studied as completely and is the main subject of this paper. 
It was first analyzed by \cite{golubov12} for one-dimensional stone walls standing on the surface of an asteroid.
Later it was computed for spherical stones \citep{golubov14} and for stones of more complicated realistic shapes \citep{sevecek15}.

An important limitation in all these works was the assumption of zero (or 180$^\circ$) obliquity of the asteroid.
This limitation raises the question of how well applicable the TYORP effect is to real asteroids
whose obliquity is not exactly zero.
Also, as the NYORP effect causes an asteroid's obliquity to migrate, this limitation also restricts the analysis of TYORP for predicting the rotational state evolution of an asteroid simultaneously acted on by NYORP and TYORP.

In this article we study TYORP dependence on obliquity.
We use a physical model, similar to \cite{golubov14},
and numeric methods, similar to \cite{sevecek15}.
Our physical model and simulation methods are explained in Section~\ref{model}.
The results of the simulations are presented and discussed in Section~\ref{results}.

\section{Methods}
\label{model}

\subsection{Physical model}

We consider spherical boulders with thermal parameter $\Theta$,
radius $R$, and at latitude $\psi$ from the equatorial plane.
The stones are half-buried in regolith,
and are situated far enough away from each other
so that any mutual shadowing or self-illumination can be neglected.
This model is similar to \cite{golubov14} (assuming $h=0$, $a=\infty$, see Table 1 therein). 

%The model has following parameters: 
%the thermal parameter $\Theta$, 
%the dimensionless boulder radius $R/L$, 
%the obliquity $\epsilon$,
%the latitude $\psi$
%and the angle $\upsilon$ between the equinox and the Sun.

By choosing to model spherical boulders we have east-west symmetry and do not have to deal with geometrical components 
of the torque that have to be ``filtered out'' by averaging over different orientations of a boulder \citep[see Section 3.3 in][]{sevecek15}.
The spherical shape is acceptable as \cite{sevecek15} showed that the results 
between spherical and irregular boulders are not dramatically different.
% In the previous paper, we used a simple $x=cos \phi$, $y=sin\phi$. We then studied the dependence on longitude,  the coordinates got more complicated. Here, we assume generaly non-zero obliquity. This brings two more parameters into the problem: the obliquity $\epsilon$ and the angle ... $\upsilon$. Coordinates ...

The torque exerted by a boulder varies as the asteroid rotates, making it necessary to average the torque over the rotation period to 
determine the secular effect.
However, for an asteroid with non-zero obliquity $\epsilon$, 
the torque also varies during the revolution around the Sun.
Because of that, it is also necessary to average the torque over 
the orbital period as well. We assume a zero eccentricity for the heliocentric orbit.
%Therefore, we evaluate the torque on several points around the orbit 
%-nd then compute the mean value. 
%We generalize the problem to non-zero obliquity but we still asume zero eccentricity. The asteroid thus revolves around the Sun with constant angular frequency. 

To obtain the torque exerted by the boulder, we first need to solve
the three-dimensional heat diffusion equation in the boulder and its surroundings,
\begin{equation}
    \nabla\cdot (K\nabla u) - \rho C \partial_t u = 0\,.
    \label{eq: }
\end{equation} 
Here $u$ is the temperature, 
$K$ is the thermal conductivity, 
$C$ is the heat capacity, and 
$\rho$ is the density.
We consider generally different values of thermal conductivity
for the boulder and the surrounding regolith.
The boundary condition on the surface is:
\begin{equation}
    K\partial_n u + \epsilon \sigma u^4 = (1-A)\Phi_\odot \mu \vec s \cdot \vec n\,,
    \label{eq: }
\end{equation}
where $\epsilon$ is the infrared emissivity,
$\sigma$~is the Stefan--Boltzmann constant,
$A$~is the hemispherical albedo,
$\Phi_\odot$~is the solar constant (at the location of the asteroid),
$\mu$~is the shadowing function,
$\vec s$~is the vector towards the Sun
and $\vec n$~is the local outward normal.

To reduce the number of independent parameters,
we perform non-dimensionalization similarly to \cite{sevecek15}.
As a unit length we choose the diurnal thermal skin depth:
\begin{equation}
    L \equiv \sqrt{\frac{2K}{\omega \rho C}}
    \label{eq: }
\end{equation}
and we adopt the following definition of the thermal parameter:
\begin{equation}
    \Theta \equiv \frac{\sqrt{K\omega \rho C}}{4\sqrt{2}\pi^{-\frac{3}{4}}\epsilon\sigma u_\star^3} \,.
    \label{eq: }
\end{equation}
Using these quantities, the TYORP torque exerted by the boulder has four independent parameters:
the dimensionless radius~$R/L$ of the boulder, the thermal parameter~$\Theta$, the latitude~$\psi$ and the obliquity~$\epsilon$.

Since we assume non-zero obliquity~$\epsilon$ and latitude~$\psi$, 
the coordinates of the Sun in a local (topocentric) frame are 
\begin{align}
     x &= -\cos\upsilon\sin\phi + \cos\epsilon\sin\upsilon\cos\phi \,,\\
    y &= -(\cos\upsilon \cos \phi + \cos\epsilon\sin\upsilon\sin \phi)\sin\psi + 
\sin\epsilon\sin \upsilon\cos \psi \,,\\
    z &= (\cos\upsilon \cos \phi + \cos\epsilon\sin\upsilon\sin \phi)\cos\psi + \sin\epsilon\sin \upsilon \sin \psi \,,
%       \label{eq: }
\end{align}
where $\upsilon$ is the angle between the point of the equinox and 
the direction towards the Sun.
(See Appendix \ref{sec:coors} for a detailed derivation.)
In these coordinates, the outward normal is $\vec n =(0,0,1)$, and 
the local meridian is aligned with $y$-axis.

\subsection{Numeric methods}

We use a numeric code similar to the one used by \cite{sevecek15}.
The heat diffusion equation is solved using the 
finite element discretization in space.
The temperature is approximated by a linear combination of prescribed basis functions
\begin{equation}
    u(\vec r, t) \simeq \sum_{j=1}^M u_j(t) N_j(\vec r)\,.
    \label{eq: }
\end{equation}
%Although the heat diffusion equation will not be satisfied pointwise,
As for the temporal derivative, we use an implicit Euler scheme with constant time-step.
We derive a weak formulation of the problem \citep[see section 2.2 of][]{sevecek15}
and we solve the resulting system of linear algebraic equations using the \texttt{FreeFem++} code \citep{hecht12}. 
The computational domain is prepared from a simple geometry 
(a spherical boulder in a regolith block) using the \texttt{tetgen} code \citep{si06}.

%The method for evaluating the magnitude of the effect is essentially similar to the one used in Sevecek et al. (2015).
Once we obtain the temperature distribution, %we compute the mean dimensionless pressure $\mdp$.
we express the magnitude of the torque using the day-averaged dimensionless pressure $\langle\Pi\rangle$:
\begin{equation}
    \langle\Pi\rangle \equiv  %\frac{1}{2\pi}
    -\frac{2}{3} \frac{1}{P} \frac{1}{\pi R^2 } %\int\limits_0^{2\pi}%
    \int\limits_0^P \int\limits_{\partial\Omega} \frac{u^4}{u_\star^4}n_x\,\d S\d t \,,%\,\d\upsilon\,,
\end{equation}
where $P$ is the rotational period,
$R$ is the radius of the spherical boulder,  
$u$ is the temperature at given point on the surface,
$u_\star \equiv [(1-A)\Phi_\odot / (\sigma\epsilon)]^{\frac{1}{4}}$ is the subsolar temperature, and 
$n_x$ is the $x$-component of the (outward) normal.
The negative sign expresses that the pressure acts against the normal,
while the coefficient 2/3 comes from integrating over all possible directions of the emitted light assuming Lambert's emission indicatrix.
%In other words, we integrate the recoil force (expressed as suitable dimensionless quantity) over the surface of the boulder and average the result over 
%the rotational and orbital period. 

Lastly, we need to account for variations of the torque during its revolution around the Sun.
We thus define the year-averaged dimensionless pressure $\mdp$ (or simply the \emph{mean} dimensionless pressure) 
\begin{equation}
    \mdp \equiv \frac{1}{2\pi} \int\limits_0^{2\pi} \langle\Pi\rangle(\upsilon)\,\d\upsilon\,.
      \label{eq: }
\end{equation}
Note that we assume zero eccentricity and thus can integrate over the angle $\upsilon$ rather
than the time.
%We denote the dimensionless pressure with double 
%averaging symbol to differentiate it from the day-averaged pressure.

In practice, we evaluate $\langle\Pi\rangle$ at a finite number of points around the 
orbit and average these values to estimate the integral.

For our purposes, the mean dimensionless pressure $\mdp$ is the final result. 
Previous works \citep{golubov12, golubov14, sevecek15} discussed the total torque 
boulders will exert on an asteroid and the corresponding change of the angular frequency; 
in this paper, however, we are focusing mainly on the dependence of the torque 
on different parameters rather than on its absolute magnitude, and thus
$\mdp$ is a useful quantity for this purpose.

The parametric space is quite extended and cannot be studied thoroughly
with available computational resources. 
We thus perform sections through parametric space by varying one parameter and keeping all other parameters constant at 
reasonable, physically relevant values.

To reach a stationary solution we need to evolve the system for several rotational periods.
For low values of $\psi$ and $\epsilon$,
the convergence is fast, and only 4-5 periods are required. At high latitudes and obliquities, however, the analytical solution is not a good approximation and up to 50 periods are needed to achieve the same accuracy.
Runs requiring the most periods are the ones where the boulder is in the polar night
and the Sun does not appear above the horizon during the rotational period at all.
However, it is obvious that the pressure $\langle \Pi\rangle$ will
be zero in these cases. Therefore, to save computation time, 
the boulder is assumed to be in polar night and the run is skipped altogether, if the following condition is met:
\begin{equation}
    \tan \psi\sin \upsilon\sin\epsilon  \leq -\sqrt{\cos^2 \upsilon + \cos^2 \epsilon\sin^2\upsilon} \,.
    \label{eq:polar_night}
\end{equation}
We confirmed on a test run that the code will return a negligible pressure of $\langle\Pi\rangle \simeq 10^{-7}$ if the condition (\ref{eq:polar_night}) is met, even though it takes about 300 periods.

%\begin{align}
%    x &= (\cos\upsilon \cos \phi + \cos\epsilon\sin\upsilon\sin \phi)\cos\psi + \sin\epsilon\sin \upsilon \sin \psi \,, \\
%    y &= -\cos\upsilon\sin\phi + \cos\epsilon\sin\upsilon\cos\phi \,\\
%    z &= -(\cos\upsilon \cos \phi + \cos\epsilon\sin\upsilon\sin \phi)\sin\psi + 
%\sin\epsilon\sin \upsilon\cos \psi \,. 
%       \label{eq: }
%\end{align}

\section{Results}
\label{results}
\begin{figure*}
    % GNUPLOT: LaTeX picture with Postscript
\begingroup
  \makeatletter
  \providecommand\color[2][]{%
    \GenericError{(gnuplot) \space\space\space\@spaces}{%
      Package color not loaded in conjunction with
      terminal option `colourtext'%
    }{See the gnuplot documentation for explanation.%
    }{Either use 'blacktext' in gnuplot or load the package
      color.sty in LaTeX.}%
    \renewcommand\color[2][]{}%
  }%
  \providecommand\includegraphics[2][]{%
    \GenericError{(gnuplot) \space\space\space\@spaces}{%
      Package graphicx or graphics not loaded%
    }{See the gnuplot documentation for explanation.%
    }{The gnuplot epslatex terminal needs graphicx.sty or graphics.sty.}%
    \renewcommand\includegraphics[2][]{}%
  }%
  \providecommand\rotatebox[2]{#2}%
  \@ifundefined{ifGPcolor}{%
    \newif\ifGPcolor
    \GPcolortrue
  }{}%
  \@ifundefined{ifGPblacktext}{%
    \newif\ifGPblacktext
    \GPblacktextfalse
  }{}%
  % define a \g@addto@macro without @ in the name:
  \let\gplgaddtomacro\g@addto@macro
  % define empty templates for all commands taking text:
  \gdef\gplbacktext{}%
  \gdef\gplfronttext{}%
  \makeatother
  \ifGPblacktext
    % no textcolor at all
    \def\colorrgb#1{}%
    \def\colorgray#1{}%
  \else
    % gray or color?
    \ifGPcolor
      \def\colorrgb#1{\color[rgb]{#1}}%
      \def\colorgray#1{\color[gray]{#1}}%
      \expandafter\def\csname LTw\endcsname{\color{white}}%
      \expandafter\def\csname LTb\endcsname{\color{black}}%
      \expandafter\def\csname LTa\endcsname{\color{black}}%
      \expandafter\def\csname LT0\endcsname{\color[rgb]{1,0,0}}%
      \expandafter\def\csname LT1\endcsname{\color[rgb]{0,1,0}}%
      \expandafter\def\csname LT2\endcsname{\color[rgb]{0,0,1}}%
      \expandafter\def\csname LT3\endcsname{\color[rgb]{1,0,1}}%
      \expandafter\def\csname LT4\endcsname{\color[rgb]{0,1,1}}%
      \expandafter\def\csname LT5\endcsname{\color[rgb]{1,1,0}}%
      \expandafter\def\csname LT6\endcsname{\color[rgb]{0,0,0}}%
      \expandafter\def\csname LT7\endcsname{\color[rgb]{1,0.3,0}}%
      \expandafter\def\csname LT8\endcsname{\color[rgb]{0.5,0.5,0.5}}%
    \else
      % gray
      \def\colorrgb#1{\color{black}}%
      \def\colorgray#1{\color[gray]{#1}}%
      \expandafter\def\csname LTw\endcsname{\color{white}}%
      \expandafter\def\csname LTb\endcsname{\color{black}}%
      \expandafter\def\csname LTa\endcsname{\color{black}}%
      \expandafter\def\csname LT0\endcsname{\color{black}}%
      \expandafter\def\csname LT1\endcsname{\color{black}}%
      \expandafter\def\csname LT2\endcsname{\color{black}}%
      \expandafter\def\csname LT3\endcsname{\color{black}}%
      \expandafter\def\csname LT4\endcsname{\color{black}}%
      \expandafter\def\csname LT5\endcsname{\color{black}}%
      \expandafter\def\csname LT6\endcsname{\color{black}}%
      \expandafter\def\csname LT7\endcsname{\color{black}}%
      \expandafter\def\csname LT8\endcsname{\color{black}}%
    \fi
  \fi
    \setlength{\unitlength}{0.0500bp}%
    \ifx\gptboxheight\undefined%
      \newlength{\gptboxheight}%
      \newlength{\gptboxwidth}%
      \newsavebox{\gptboxtext}%
    \fi%
    \setlength{\fboxrule}{0.5pt}%
    \setlength{\fboxsep}{1pt}%
\begin{picture}(12960.00,2880.00)%
    \gplgaddtomacro\gplbacktext{%
      \csname LTb\endcsname%
      \put(814,704){\makebox(0,0)[r]{\strut{}$0$}}%
      \put(814,916){\makebox(0,0)[r]{\strut{}$0.5$}}%
      \put(814,1129){\makebox(0,0)[r]{\strut{}$1$}}%
      \put(814,1341){\makebox(0,0)[r]{\strut{}$1.5$}}%
      \put(814,1553){\makebox(0,0)[r]{\strut{}$2$}}%
      \put(814,1766){\makebox(0,0)[r]{\strut{}$2.5$}}%
      \put(814,1978){\makebox(0,0)[r]{\strut{}$3$}}%
      \put(814,2190){\makebox(0,0)[r]{\strut{}$3.5$}}%
      \put(814,2403){\makebox(0,0)[r]{\strut{}$4$}}%
      \put(814,2615){\makebox(0,0)[r]{\strut{}$4.5$}}%
      \put(946,484){\makebox(0,0){\strut{}$0$}}%
      \put(1442,484){\makebox(0,0){\strut{}$60$}}%
      \put(1938,484){\makebox(0,0){\strut{}$120$}}%
      \put(2435,484){\makebox(0,0){\strut{}$180$}}%
      \put(2931,484){\makebox(0,0){\strut{}$240$}}%
      \put(3427,484){\makebox(0,0){\strut{}$300$}}%
      \put(3923,484){\makebox(0,0){\strut{}$360$}}%
      \put(2186,916){\makebox(0,0)[l]{\strut{}$\psi = 0^\circ$}}%
    }%
    \gplgaddtomacro\gplfronttext{%
      \csname LTb\endcsname%
      \put(176,1659){\rotatebox{-270}{\makebox(0,0){\strut{}$\langle\Pi\rangle \cdot 10^{3}$}}}%
      \put(2434,154){\makebox(0,0){\strut{}$\upsilon$ [$^\circ$]}}%
    }%
    \gplgaddtomacro\gplbacktext{%
      \csname LTb\endcsname%
      \put(3922,484){\makebox(0,0){\strut{}}}%
      \put(4422,484){\makebox(0,0){\strut{}60}}%
      \put(4921,484){\makebox(0,0){\strut{}120}}%
      \put(5421,484){\makebox(0,0){\strut{}180}}%
      \put(5921,484){\makebox(0,0){\strut{}240}}%
      \put(6420,484){\makebox(0,0){\strut{}300}}%
      \put(6920,484){\makebox(0,0){\strut{}360}}%
      \put(5171,916){\makebox(0,0)[l]{\strut{}$\psi = 40^\circ$}}%
    }%
    \gplgaddtomacro\gplfronttext{%
      \csname LTb\endcsname%
      \put(5421,154){\makebox(0,0){\strut{}$\upsilon$ [$^\circ$]}}%
    }%
    \gplgaddtomacro\gplbacktext{%
      \csname LTb\endcsname%
      \put(6920,484){\makebox(0,0){\strut{}}}%
      \put(7376,484){\makebox(0,0){\strut{}60}}%
      \put(7831,484){\makebox(0,0){\strut{}120}}%
      \put(8287,484){\makebox(0,0){\strut{}180}}%
      \put(8742,484){\makebox(0,0){\strut{}240}}%
      \put(9198,484){\makebox(0,0){\strut{}300}}%
      \put(9653,484){\makebox(0,0){\strut{}360}}%
      \put(9785,704){\makebox(0,0)[l]{\strut{}}}%
      \put(9785,916){\makebox(0,0)[l]{\strut{}}}%
      \put(9785,1129){\makebox(0,0)[l]{\strut{}}}%
      \put(9785,1341){\makebox(0,0)[l]{\strut{}}}%
      \put(9785,1553){\makebox(0,0)[l]{\strut{}}}%
      \put(9785,1766){\makebox(0,0)[l]{\strut{}}}%
      \put(9785,1978){\makebox(0,0)[l]{\strut{}}}%
      \put(9785,2190){\makebox(0,0)[l]{\strut{}}}%
      \put(9785,2403){\makebox(0,0)[l]{\strut{}}}%
      \put(9785,2615){\makebox(0,0)[l]{\strut{}}}%
      \put(8059,916){\makebox(0,0)[l]{\strut{}$\psi = 80^\circ$}}%
    }%
    \gplgaddtomacro\gplfronttext{%
      \csname LTb\endcsname%
      \put(8286,154){\makebox(0,0){\strut{}$\upsilon$ [$^\circ$]}}%
      \csname LTb\endcsname%
      \put(8666,2442){\makebox(0,0)[r]{\strut{}$\epsilon = 0^\circ$}}%
      \csname LTb\endcsname%
      \put(8666,2222){\makebox(0,0)[r]{\strut{}$\epsilon = 40^\circ$}}%
      \csname LTb\endcsname%
      \put(8666,2002){\makebox(0,0)[r]{\strut{}$\epsilon = 80^\circ$}}%
    }%
    \gplbacktext
    \put(0,0){\includegraphics{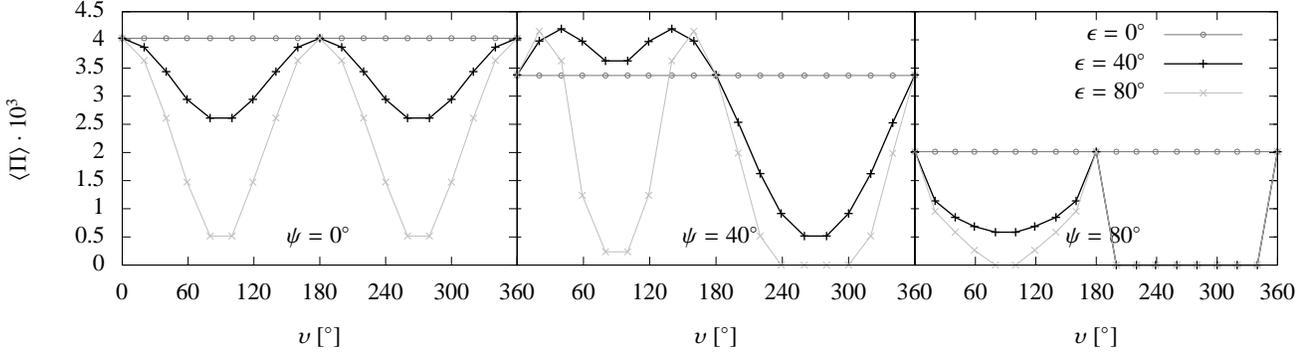}}%
    \gplfronttext
  \end{picture}%
\endgroup    \caption{Variations of the day-averaged dimensionless pressure $\langle\Pi\rangle$ during the revolution around the Sun. 
    Various values of latitude $\psi$ and obliquity $\epsilon$ are plotted.}
\label{fig:orbital_variations}
\end{figure*}
\begin{figure*}
% GNUPLOT: LaTeX picture with Postscript
\begingroup
  \makeatletter
  \providecommand\color[2][]{%
    \GenericError{(gnuplot) \space\space\space\@spaces}{%
      Package color not loaded in conjunction with
      terminal option `colourtext'%
    }{See the gnuplot documentation for explanation.%
    }{Either use 'blacktext' in gnuplot or load the package
      color.sty in LaTeX.}%
    \renewcommand\color[2][]{}%
  }%
  \providecommand\includegraphics[2][]{%
    \GenericError{(gnuplot) \space\space\space\@spaces}{%
      Package graphicx or graphics not loaded%
    }{See the gnuplot documentation for explanation.%
    }{The gnuplot epslatex terminal needs graphicx.sty or graphics.sty.}%
    \renewcommand\includegraphics[2][]{}%
  }%
  \providecommand\rotatebox[2]{#2}%
  \@ifundefined{ifGPcolor}{%
    \newif\ifGPcolor
    \GPcolortrue
  }{}%
  \@ifundefined{ifGPblacktext}{%
    \newif\ifGPblacktext
    \GPblacktextfalse
  }{}%
  % define a \g@addto@macro without @ in the name:
  \let\gplgaddtomacro\g@addto@macro
  % define empty templates for all commands taking text:
  \gdef\gplbacktext{}%
  \gdef\gplfronttext{}%
  \makeatother
  \ifGPblacktext
    % no textcolor at all
    \def\colorrgb#1{}%
    \def\colorgray#1{}%
  \else
    % gray or color?
    \ifGPcolor
      \def\colorrgb#1{\color[rgb]{#1}}%
      \def\colorgray#1{\color[gray]{#1}}%
      \expandafter\def\csname LTw\endcsname{\color{white}}%
      \expandafter\def\csname LTb\endcsname{\color{black}}%
      \expandafter\def\csname LTa\endcsname{\color{black}}%
      \expandafter\def\csname LT0\endcsname{\color[rgb]{1,0,0}}%
      \expandafter\def\csname LT1\endcsname{\color[rgb]{0,1,0}}%
      \expandafter\def\csname LT2\endcsname{\color[rgb]{0,0,1}}%
      \expandafter\def\csname LT3\endcsname{\color[rgb]{1,0,1}}%
      \expandafter\def\csname LT4\endcsname{\color[rgb]{0,1,1}}%
      \expandafter\def\csname LT5\endcsname{\color[rgb]{1,1,0}}%
      \expandafter\def\csname LT6\endcsname{\color[rgb]{0,0,0}}%
      \expandafter\def\csname LT7\endcsname{\color[rgb]{1,0.3,0}}%
      \expandafter\def\csname LT8\endcsname{\color[rgb]{0.5,0.5,0.5}}%
    \else
      % gray
      \def\colorrgb#1{\color{black}}%
      \def\colorgray#1{\color[gray]{#1}}%
      \expandafter\def\csname LTw\endcsname{\color{white}}%
      \expandafter\def\csname LTb\endcsname{\color{black}}%
      \expandafter\def\csname LTa\endcsname{\color{black}}%
      \expandafter\def\csname LT0\endcsname{\color{black}}%
      \expandafter\def\csname LT1\endcsname{\color{black}}%
      \expandafter\def\csname LT2\endcsname{\color{black}}%
      \expandafter\def\csname LT3\endcsname{\color{black}}%
      \expandafter\def\csname LT4\endcsname{\color{black}}%
      \expandafter\def\csname LT5\endcsname{\color{black}}%
      \expandafter\def\csname LT6\endcsname{\color{black}}%
      \expandafter\def\csname LT7\endcsname{\color{black}}%
      \expandafter\def\csname LT8\endcsname{\color{black}}%
    \fi
  \fi
    \setlength{\unitlength}{0.0500bp}%
    \ifx\gptboxheight\undefined%
      \newlength{\gptboxheight}%
      \newlength{\gptboxwidth}%
      \newsavebox{\gptboxtext}%
    \fi%
    \setlength{\fboxrule}{0.5pt}%
    \setlength{\fboxsep}{1pt}%
\begin{picture}(5040.00,2880.00)%
    \gplgaddtomacro\gplbacktext{%
      \csname LTb\endcsname%
      \put(682,704){\makebox(0,0)[r]{\strut{}$0$}}%
      \put(682,916){\makebox(0,0)[r]{\strut{}$0.5$}}%
      \put(682,1129){\makebox(0,0)[r]{\strut{}$1$}}%
      \put(682,1341){\makebox(0,0)[r]{\strut{}$1.5$}}%
      \put(682,1553){\makebox(0,0)[r]{\strut{}$2$}}%
      \put(682,1766){\makebox(0,0)[r]{\strut{}$2.5$}}%
      \put(682,1978){\makebox(0,0)[r]{\strut{}$3$}}%
      \put(682,2190){\makebox(0,0)[r]{\strut{}$3.5$}}%
      \put(682,2403){\makebox(0,0)[r]{\strut{}$4$}}%
      \put(682,2615){\makebox(0,0)[r]{\strut{}$4.5$}}%
      \put(814,484){\makebox(0,0){\strut{}0.1}}%
      \put(1757,484){\makebox(0,0){\strut{}0.3}}%
      \put(2791,484){\makebox(0,0){\strut{}1}}%
      \put(3734,484){\makebox(0,0){\strut{}3}}%
    }%
    \gplgaddtomacro\gplfronttext{%
      \csname LTb\endcsname%
      \put(176,1659){\rotatebox{-270}{\makebox(0,0){\strut{}$\mdp \cdot 10^{3} $}}}%
      \put(2728,154){\makebox(0,0){\strut{}$R/L$}}%
      \csname LTb\endcsname%
      \put(1606,2442){\makebox(0,0)[r]{\strut{}$\epsilon=0^\circ$}}%
      \csname LTb\endcsname%
      \put(1606,2222){\makebox(0,0)[r]{\strut{}$\epsilon=45^\circ$}}%
      \csname LTb\endcsname%
      \put(1606,2002){\makebox(0,0)[r]{\strut{}$\epsilon=90^\circ$}}%
    }%
    \gplbacktext
    \put(0,0){\includegraphics{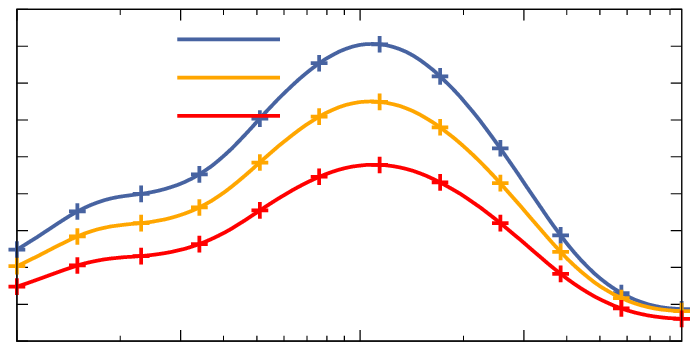}}%
    \gplfronttext
  \end{picture}%
\endgroup
% GNUPLOT: LaTeX picture with Postscript
\begingroup
  \makeatletter
  \providecommand\color[2][]{%
    \GenericError{(gnuplot) \space\space\space\@spaces}{%
      Package color not loaded in conjunction with
      terminal option `colourtext'%
    }{See the gnuplot documentation for explanation.%
    }{Either use 'blacktext' in gnuplot or load the package
      color.sty in LaTeX.}%
    \renewcommand\color[2][]{}%
  }%
  \providecommand\includegraphics[2][]{%
    \GenericError{(gnuplot) \space\space\space\@spaces}{%
      Package graphicx or graphics not loaded%
    }{See the gnuplot documentation for explanation.%
    }{The gnuplot epslatex terminal needs graphicx.sty or graphics.sty.}%
    \renewcommand\includegraphics[2][]{}%
  }%
  \providecommand\rotatebox[2]{#2}%
  \@ifundefined{ifGPcolor}{%
    \newif\ifGPcolor
    \GPcolortrue
  }{}%
  \@ifundefined{ifGPblacktext}{%
    \newif\ifGPblacktext
    \GPblacktextfalse
  }{}%
  % define a \g@addto@macro without @ in the name:
  \let\gplgaddtomacro\g@addto@macro
  % define empty templates for all commands taking text:
  \gdef\gplbacktext{}%
  \gdef\gplfronttext{}%
  \makeatother
  \ifGPblacktext
    % no textcolor at all
    \def\colorrgb#1{}%
    \def\colorgray#1{}%
  \else
    % gray or color?
    \ifGPcolor
      \def\colorrgb#1{\color[rgb]{#1}}%
      \def\colorgray#1{\color[gray]{#1}}%
      \expandafter\def\csname LTw\endcsname{\color{white}}%
      \expandafter\def\csname LTb\endcsname{\color{black}}%
      \expandafter\def\csname LTa\endcsname{\color{black}}%
      \expandafter\def\csname LT0\endcsname{\color[rgb]{1,0,0}}%
      \expandafter\def\csname LT1\endcsname{\color[rgb]{0,1,0}}%
      \expandafter\def\csname LT2\endcsname{\color[rgb]{0,0,1}}%
      \expandafter\def\csname LT3\endcsname{\color[rgb]{1,0,1}}%
      \expandafter\def\csname LT4\endcsname{\color[rgb]{0,1,1}}%
      \expandafter\def\csname LT5\endcsname{\color[rgb]{1,1,0}}%
      \expandafter\def\csname LT6\endcsname{\color[rgb]{0,0,0}}%
      \expandafter\def\csname LT7\endcsname{\color[rgb]{1,0.3,0}}%
      \expandafter\def\csname LT8\endcsname{\color[rgb]{0.5,0.5,0.5}}%
    \else
      % gray
      \def\colorrgb#1{\color{black}}%
      \def\colorgray#1{\color[gray]{#1}}%
      \expandafter\def\csname LTw\endcsname{\color{white}}%
      \expandafter\def\csname LTb\endcsname{\color{black}}%
      \expandafter\def\csname LTa\endcsname{\color{black}}%
      \expandafter\def\csname LT0\endcsname{\color{black}}%
      \expandafter\def\csname LT1\endcsname{\color{black}}%
      \expandafter\def\csname LT2\endcsname{\color{black}}%
      \expandafter\def\csname LT3\endcsname{\color{black}}%
      \expandafter\def\csname LT4\endcsname{\color{black}}%
      \expandafter\def\csname LT5\endcsname{\color{black}}%
      \expandafter\def\csname LT6\endcsname{\color{black}}%
      \expandafter\def\csname LT7\endcsname{\color{black}}%
      \expandafter\def\csname LT8\endcsname{\color{black}}%
    \fi
  \fi
    \setlength{\unitlength}{0.0500bp}%
    \ifx\gptboxheight\undefined%
      \newlength{\gptboxheight}%
      \newlength{\gptboxwidth}%
      \newsavebox{\gptboxtext}%
    \fi%
    \setlength{\fboxrule}{0.5pt}%
    \setlength{\fboxsep}{1pt}%
\begin{picture}(5040.00,2880.00)%
    \gplgaddtomacro\gplbacktext{%
      \csname LTb\endcsname%
      \put(682,704){\makebox(0,0)[r]{\strut{}$0$}}%
      \put(682,916){\makebox(0,0)[r]{\strut{}$0.5$}}%
      \put(682,1129){\makebox(0,0)[r]{\strut{}$1$}}%
      \put(682,1341){\makebox(0,0)[r]{\strut{}$1.5$}}%
      \put(682,1553){\makebox(0,0)[r]{\strut{}$2$}}%
      \put(682,1766){\makebox(0,0)[r]{\strut{}$2.5$}}%
      \put(682,1978){\makebox(0,0)[r]{\strut{}$3$}}%
      \put(682,2190){\makebox(0,0)[r]{\strut{}$3.5$}}%
      \put(682,2403){\makebox(0,0)[r]{\strut{}$4$}}%
      \put(682,2615){\makebox(0,0)[r]{\strut{}$4.5$}}%
      \put(814,484){\makebox(0,0){\strut{}0.1}}%
      \put(1757,484){\makebox(0,0){\strut{}0.3}}%
      \put(2791,484){\makebox(0,0){\strut{}1}}%
      \put(3734,484){\makebox(0,0){\strut{}3}}%
    }%
    \gplgaddtomacro\gplfronttext{%
      \csname LTb\endcsname%
      \put(176,1659){\rotatebox{-270}{\makebox(0,0){\strut{}$\mdp \cdot 10^{3} $}}}%
      \put(2728,154){\makebox(0,0){\strut{}$\Theta$}}%
      \csname LTb\endcsname%
      \put(1606,1317){\makebox(0,0)[r]{\strut{}$\epsilon=0^\circ$}}%
      \csname LTb\endcsname%
      \put(1606,1097){\makebox(0,0)[r]{\strut{}$\epsilon=45^\circ$}}%
      \csname LTb\endcsname%
      \put(1606,877){\makebox(0,0)[r]{\strut{}$\epsilon=90^\circ$}}%
    }%
    \gplbacktext
    \put(0,0){\includegraphics{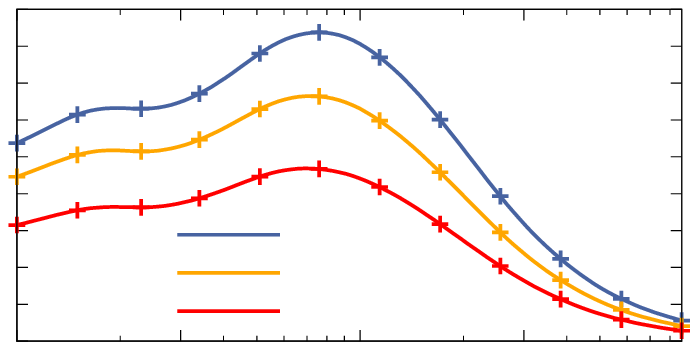}}%
    \gplfronttext
  \end{picture}%
\endgroup \par
% GNUPLOT: LaTeX picture with Postscript
\begingroup
  \makeatletter
  \providecommand\color[2][]{%
    \GenericError{(gnuplot) \space\space\space\@spaces}{%
      Package color not loaded in conjunction with
      terminal option `colourtext'%
    }{See the gnuplot documentation for explanation.%
    }{Either use 'blacktext' in gnuplot or load the package
      color.sty in LaTeX.}%
    \renewcommand\color[2][]{}%
  }%
  \providecommand\includegraphics[2][]{%
    \GenericError{(gnuplot) \space\space\space\@spaces}{%
      Package graphicx or graphics not loaded%
    }{See the gnuplot documentation for explanation.%
    }{The gnuplot epslatex terminal needs graphicx.sty or graphics.sty.}%
    \renewcommand\includegraphics[2][]{}%
  }%
  \providecommand\rotatebox[2]{#2}%
  \@ifundefined{ifGPcolor}{%
    \newif\ifGPcolor
    \GPcolortrue
  }{}%
  \@ifundefined{ifGPblacktext}{%
    \newif\ifGPblacktext
    \GPblacktextfalse
  }{}%
  % define a \g@addto@macro without @ in the name:
  \let\gplgaddtomacro\g@addto@macro
  % define empty templates for all commands taking text:
  \gdef\gplbacktext{}%
  \gdef\gplfronttext{}%
  \makeatother
  \ifGPblacktext
    % no textcolor at all
    \def\colorrgb#1{}%
    \def\colorgray#1{}%
  \else
    % gray or color?
    \ifGPcolor
      \def\colorrgb#1{\color[rgb]{#1}}%
      \def\colorgray#1{\color[gray]{#1}}%
      \expandafter\def\csname LTw\endcsname{\color{white}}%
      \expandafter\def\csname LTb\endcsname{\color{black}}%
      \expandafter\def\csname LTa\endcsname{\color{black}}%
      \expandafter\def\csname LT0\endcsname{\color[rgb]{1,0,0}}%
      \expandafter\def\csname LT1\endcsname{\color[rgb]{0,1,0}}%
      \expandafter\def\csname LT2\endcsname{\color[rgb]{0,0,1}}%
      \expandafter\def\csname LT3\endcsname{\color[rgb]{1,0,1}}%
      \expandafter\def\csname LT4\endcsname{\color[rgb]{0,1,1}}%
      \expandafter\def\csname LT5\endcsname{\color[rgb]{1,1,0}}%
      \expandafter\def\csname LT6\endcsname{\color[rgb]{0,0,0}}%
      \expandafter\def\csname LT7\endcsname{\color[rgb]{1,0.3,0}}%
      \expandafter\def\csname LT8\endcsname{\color[rgb]{0.5,0.5,0.5}}%
    \else
      % gray
      \def\colorrgb#1{\color{black}}%
      \def\colorgray#1{\color[gray]{#1}}%
      \expandafter\def\csname LTw\endcsname{\color{white}}%
      \expandafter\def\csname LTb\endcsname{\color{black}}%
      \expandafter\def\csname LTa\endcsname{\color{black}}%
      \expandafter\def\csname LT0\endcsname{\color{black}}%
      \expandafter\def\csname LT1\endcsname{\color{black}}%
      \expandafter\def\csname LT2\endcsname{\color{black}}%
      \expandafter\def\csname LT3\endcsname{\color{black}}%
      \expandafter\def\csname LT4\endcsname{\color{black}}%
      \expandafter\def\csname LT5\endcsname{\color{black}}%
      \expandafter\def\csname LT6\endcsname{\color{black}}%
      \expandafter\def\csname LT7\endcsname{\color{black}}%
      \expandafter\def\csname LT8\endcsname{\color{black}}%
    \fi
  \fi
    \setlength{\unitlength}{0.0500bp}%
    \ifx\gptboxheight\undefined%
      \newlength{\gptboxheight}%
      \newlength{\gptboxwidth}%
      \newsavebox{\gptboxtext}%
    \fi%
    \setlength{\fboxrule}{0.5pt}%
    \setlength{\fboxsep}{1pt}%
\begin{picture}(5040.00,2880.00)%
    \gplgaddtomacro\gplbacktext{%
      \csname LTb\endcsname%
      \put(682,704){\makebox(0,0)[r]{\strut{}$0$}}%
      \put(682,916){\makebox(0,0)[r]{\strut{}$0.5$}}%
      \put(682,1129){\makebox(0,0)[r]{\strut{}$1$}}%
      \put(682,1341){\makebox(0,0)[r]{\strut{}$1.5$}}%
      \put(682,1553){\makebox(0,0)[r]{\strut{}$2$}}%
      \put(682,1766){\makebox(0,0)[r]{\strut{}$2.5$}}%
      \put(682,1978){\makebox(0,0)[r]{\strut{}$3$}}%
      \put(682,2190){\makebox(0,0)[r]{\strut{}$3.5$}}%
      \put(682,2403){\makebox(0,0)[r]{\strut{}$4$}}%
      \put(682,2615){\makebox(0,0)[r]{\strut{}$4.5$}}%
      \put(814,484){\makebox(0,0){\strut{}$0$}}%
      \put(1239,484){\makebox(0,0){\strut{}$10$}}%
      \put(1665,484){\makebox(0,0){\strut{}$20$}}%
      \put(2090,484){\makebox(0,0){\strut{}$30$}}%
      \put(2516,484){\makebox(0,0){\strut{}$40$}}%
      \put(2941,484){\makebox(0,0){\strut{}$50$}}%
      \put(3367,484){\makebox(0,0){\strut{}$60$}}%
      \put(3792,484){\makebox(0,0){\strut{}$70$}}%
      \put(4218,484){\makebox(0,0){\strut{}$80$}}%
      \put(4643,484){\makebox(0,0){\strut{}$90$}}%
    }%
    \gplgaddtomacro\gplfronttext{%
      \csname LTb\endcsname%
      \put(176,1659){\rotatebox{-270}{\makebox(0,0){\strut{}$\mdp \cdot 10^{3}$}}}%
      \put(2728,154){\makebox(0,0){\strut{}$\epsilon$ [$^\circ$]}}%
      \csname LTb\endcsname%
      \put(3656,2442){\makebox(0,0)[r]{\strut{}$\psi=0^\circ$}}%
      \csname LTb\endcsname%
      \put(3656,2222){\makebox(0,0)[r]{\strut{}$\psi=40^\circ$}}%
      \csname LTb\endcsname%
      \put(3656,2002){\makebox(0,0)[r]{\strut{}$\psi=80^\circ$}}%
    }%
    \gplbacktext
    \put(0,0){\includegraphics{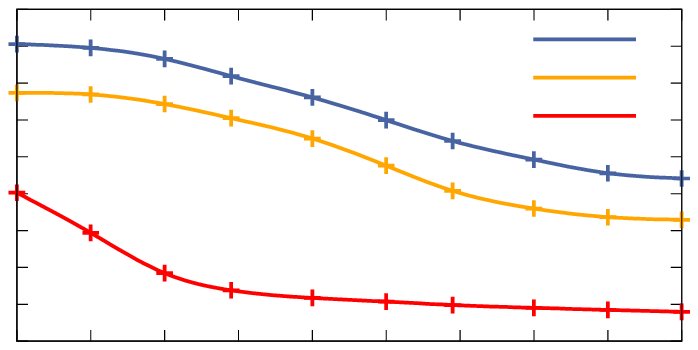}}%
    \gplfronttext
  \end{picture}%
\endgroup % GNUPLOT: LaTeX picture with Postscript
\begingroup
  \makeatletter
  \providecommand\color[2][]{%
    \GenericError{(gnuplot) \space\space\space\@spaces}{%
      Package color not loaded in conjunction with
      terminal option `colourtext'%
    }{See the gnuplot documentation for explanation.%
    }{Either use 'blacktext' in gnuplot or load the package
      color.sty in LaTeX.}%
    \renewcommand\color[2][]{}%
  }%
  \providecommand\includegraphics[2][]{%
    \GenericError{(gnuplot) \space\space\space\@spaces}{%
      Package graphicx or graphics not loaded%
    }{See the gnuplot documentation for explanation.%
    }{The gnuplot epslatex terminal needs graphicx.sty or graphics.sty.}%
    \renewcommand\includegraphics[2][]{}%
  }%
  \providecommand\rotatebox[2]{#2}%
  \@ifundefined{ifGPcolor}{%
    \newif\ifGPcolor
    \GPcolortrue
  }{}%
  \@ifundefined{ifGPblacktext}{%
    \newif\ifGPblacktext
    \GPblacktextfalse
  }{}%
  % define a \g@addto@macro without @ in the name:
  \let\gplgaddtomacro\g@addto@macro
  % define empty templates for all commands taking text:
  \gdef\gplbacktext{}%
  \gdef\gplfronttext{}%
  \makeatother
  \ifGPblacktext
    % no textcolor at all
    \def\colorrgb#1{}%
    \def\colorgray#1{}%
  \else
    % gray or color?
    \ifGPcolor
      \def\colorrgb#1{\color[rgb]{#1}}%
      \def\colorgray#1{\color[gray]{#1}}%
      \expandafter\def\csname LTw\endcsname{\color{white}}%
      \expandafter\def\csname LTb\endcsname{\color{black}}%
      \expandafter\def\csname LTa\endcsname{\color{black}}%
      \expandafter\def\csname LT0\endcsname{\color[rgb]{1,0,0}}%
      \expandafter\def\csname LT1\endcsname{\color[rgb]{0,1,0}}%
      \expandafter\def\csname LT2\endcsname{\color[rgb]{0,0,1}}%
      \expandafter\def\csname LT3\endcsname{\color[rgb]{1,0,1}}%
      \expandafter\def\csname LT4\endcsname{\color[rgb]{0,1,1}}%
      \expandafter\def\csname LT5\endcsname{\color[rgb]{1,1,0}}%
      \expandafter\def\csname LT6\endcsname{\color[rgb]{0,0,0}}%
      \expandafter\def\csname LT7\endcsname{\color[rgb]{1,0.3,0}}%
      \expandafter\def\csname LT8\endcsname{\color[rgb]{0.5,0.5,0.5}}%
    \else
      % gray
      \def\colorrgb#1{\color{black}}%
      \def\colorgray#1{\color[gray]{#1}}%
      \expandafter\def\csname LTw\endcsname{\color{white}}%
      \expandafter\def\csname LTb\endcsname{\color{black}}%
      \expandafter\def\csname LTa\endcsname{\color{black}}%
      \expandafter\def\csname LT0\endcsname{\color{black}}%
      \expandafter\def\csname LT1\endcsname{\color{black}}%
      \expandafter\def\csname LT2\endcsname{\color{black}}%
      \expandafter\def\csname LT3\endcsname{\color{black}}%
      \expandafter\def\csname LT4\endcsname{\color{black}}%
      \expandafter\def\csname LT5\endcsname{\color{black}}%
      \expandafter\def\csname LT6\endcsname{\color{black}}%
      \expandafter\def\csname LT7\endcsname{\color{black}}%
      \expandafter\def\csname LT8\endcsname{\color{black}}%
    \fi
  \fi
    \setlength{\unitlength}{0.0500bp}%
    \ifx\gptboxheight\undefined%
      \newlength{\gptboxheight}%
      \newlength{\gptboxwidth}%
      \newsavebox{\gptboxtext}%
    \fi%
    \setlength{\fboxrule}{0.5pt}%
    \setlength{\fboxsep}{1pt}%
\begin{picture}(5040.00,2880.00)%
    \gplgaddtomacro\gplbacktext{%
      \csname LTb\endcsname%
      \put(682,704){\makebox(0,0)[r]{\strut{}$0$}}%
      \put(682,916){\makebox(0,0)[r]{\strut{}$0.5$}}%
      \put(682,1129){\makebox(0,0)[r]{\strut{}$1$}}%
      \put(682,1341){\makebox(0,0)[r]{\strut{}$1.5$}}%
      \put(682,1553){\makebox(0,0)[r]{\strut{}$2$}}%
      \put(682,1766){\makebox(0,0)[r]{\strut{}$2.5$}}%
      \put(682,1978){\makebox(0,0)[r]{\strut{}$3$}}%
      \put(682,2190){\makebox(0,0)[r]{\strut{}$3.5$}}%
      \put(682,2403){\makebox(0,0)[r]{\strut{}$4$}}%
      \put(682,2615){\makebox(0,0)[r]{\strut{}$4.5$}}%
      \put(814,484){\makebox(0,0){\strut{}$0$}}%
      \put(1293,484){\makebox(0,0){\strut{}$10$}}%
      \put(1771,484){\makebox(0,0){\strut{}$20$}}%
      \put(2250,484){\makebox(0,0){\strut{}$30$}}%
      \put(2729,484){\makebox(0,0){\strut{}$40$}}%
      \put(3207,484){\makebox(0,0){\strut{}$50$}}%
      \put(3686,484){\makebox(0,0){\strut{}$60$}}%
      \put(4164,484){\makebox(0,0){\strut{}$70$}}%
      \put(4643,484){\makebox(0,0){\strut{}$80$}}%
    }%
    \gplgaddtomacro\gplfronttext{%
      \csname LTb\endcsname%
      \put(176,1659){\rotatebox{-270}{\makebox(0,0){\strut{}$\mdp \cdot 10^{3}$}}}%
      \put(2728,154){\makebox(0,0){\strut{}$\psi$ [$^\circ$]}}%
      \csname LTb\endcsname%
      \put(1606,1317){\makebox(0,0)[r]{\strut{}$\epsilon=0^\circ$}}%
      \csname LTb\endcsname%
      \put(1606,1097){\makebox(0,0)[r]{\strut{}$\epsilon=45^\circ$}}%
      \csname LTb\endcsname%
      \put(1606,877){\makebox(0,0)[r]{\strut{}$\epsilon=90^\circ$}}%
    }%
    \gplbacktext
    \put(0,0){\includegraphics{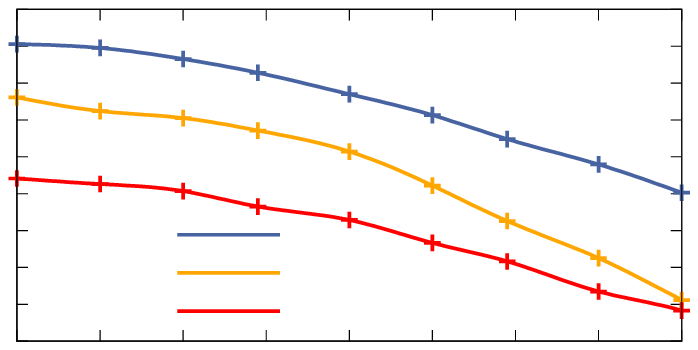}}%
    \gplfronttext
  \end{picture}%
\endgroup\caption{TYORP as a function of relevant parameters. 
The upper two panels show $\mdp$ for $\Theta=1$ and different $R/L$ (upper left panel) or $R/L=1$ and different $\Theta$ (upper left panel),
for latitude~$\psi=0^\circ$ in all cases, and for three different values of obliquity~$\epsilon$. 
The lower two panels show $\mdp$ for different values of $\epsilon$ and $\psi$, while both $\Theta$ and $R/L$ are set to 1.}
\label{fig:plots}
\end{figure*}
\begin{figure*}[h]
\centering
\begin{tikzpicture}
    \draw[] (-5, 0) circle (2.5);
    \draw[] (5, 0) circle (2.5);
    \fill[orange] (-5, 0) circle (0.5);
    \fill[orange] (5, 0) circle (0.5);
    \toprockA{-2.5}{0}
    \toprockA{-7.5}{0}
    \toprockA{-5}{2.5}
    \toprockA{-5}{-2.5}    
    \fill[lightgray] (-2.5,0) circle (1); 
    \fill[lightgray] (-7.5,0) circle (1); 
    \fill[lightgray] (-5,2.5) circle (1); 
    \fill[lightgray] (-5,-2.5) circle (1); 
    \toprockB{-2.5}{0}
    \toprockB{-7.5}{0}
    \toprockB{-5}{2.5}
    \toprockB{-5}{-2.5}    
    
    \pole{2.5}{0};
    \pole{7.5}{0};
    \pole{5}{2.5};
    \pole{5}{-2.5};
    \botrockA{2.5}{0}
    \botrockA{7.5}{0}
    \botrockA{5}{2.5}
    \botrockA{5}{-2.5}    
    \botrockB{2.5}{0}
    \botrockB{7.5}{0}
    \botrockB{5}{2.5}
    \botrockB{5}{-2.5}    

    \fill[lightgray] (2.5,0) circle (1); 
    \fill[lightgray] (7.5,0) circle (1); 
    \fill[lightgray] (5,2.5) circle (1); 
    \fill[lightgray] (5,-2.5) circle (1); 
    \spin{-2.5}{0}
    \spin{-7.5}{0}
    \spin{-5}{2.5}
    \spin{-5}{-2.5}
    %\seasons{-5}
    \seasons{5}
\end{tikzpicture}
\caption{Illustration of TYORP dependence on obliquity $\epsilon$ and latitude $\psi$.
Shown are two asteroids, with $\epsilon\approx 0^\circ$ and $\epsilon\approx 90^\circ$,
and two boulders on their surface, with $\psi\approx 0^\circ$ and $\psi\approx 90^\circ$.}
\label{fig:illustration}%
\end{figure*}

Results of our simulations are plotted in Figures \ref{fig:orbital_variations}, \ref{fig:plots}, and \ref{fig:total}.

Figure \ref{fig:orbital_variations} illustrates how the overall TYORP torque $\mdp$ is accumulated over time via contributions from different seasons.
The season-unaveraged torque $\langle\Pi\rangle$ is plotted versus $\upsilon$, the angle the asteroid has passed from the equinox.
Three different values of latitude $\psi$ are shown in different panels, and two different values of obliquity $\epsilon$ are shown with different lines.
For $\epsilon=0^\circ$ the torque $\langle\Pi\rangle$ is constant as function of $\upsilon$, 
but with increasing obliquity the seasonal distribution of $\langle\Pi\rangle$ gets more uneven.
For $\psi=0^\circ$ equinoxes produce larger contributions than solstices, which is reasonable as illumination from the East and the West is the most important for TYORP,
and this is exactly the type of illumination that is more prominent during equinoxes
than during solstices, when more illumination comes from the South (winter solstice) or from above (summer solstice).
Everywhere but the equator winter solstices give a smaller contribution than summer solstices.
Naturally, $\langle\Pi\rangle$ is zero during polar nights.

Figure \ref{fig:plots} is the central result of this article.
It illustrates how the torque $\mdp$ depends on the relevant physical parameters: thermal parameter~$\Theta$,
non-dimensional radius~$R/L$, obliquity~$\epsilon$, and latitude~$\psi$.
In each panel only two parameters vary (one is colour-coded, and the other is given on the horizontal axis), 
while the others are set to two of the following values: $\Theta=1$, $R/L=1$, $\psi=0^\circ$.
We can imagine $\mdp$ plotted above a four-dimensional parameter space,
and Figure \ref{fig:plots} presenting some sections of this space.
We only plot obliquities between 0 and 90$^\circ$, however all of the results should be symmetric about 90$^\circ$, 
with results for obliquities of 0 and 180$^\circ$ being the same.

From the upper two panels of Figure \ref{fig:plots} we see that to a reasonable accuracy $R/L$ and $\Theta$ are separable from $\epsilon$, 
so that changes in $\epsilon$ mostly cause a rescaling of the plot with the same coefficient for all $R/L$ and $\Theta$.
Parameters $\epsilon$ and $\psi$ are clearly not separable, and a change of $\psi$ causes larger relative changes in TYORP at high obliquities than at low obliquities.

Many features of the dependence of TYORP pressure $\mdp$ on obliquity $\epsilon$ and latitude $\psi$ can be understood intuitively from Figure \ref{fig:illustration}.
If an asteroid has zero obliquity, as in the left-hand panel of Figure \ref{fig:illustration}, 
then boulders A with $\psi\approx 0^\circ$ and B with $\psi\approx 90^\circ$ appear in similar conditions.
Although at noon the former is illuminated from above while the latter from the South, in the morning and in the evening
illumination conditions are exactly equivalent.
Thus it is sensible that the difference between the cases $\psi\approx 0^\circ$ and $\psi\approx 90^\circ$ are only about twofold.
A larger TYORP effect for $\psi\approx 0^\circ$ can be attributed to it having a projected area up to two times larger, thus absorbing and emitting more light.

This contrasts the result by \cite{golubov14}, who obtained $\mdp=0$ when $\psi=90^\circ$,
which happened due to the regular arrangement of stones on the surface they considered, and thus accounting for their mutual shadowing.
Here we disregard shadowing and consider only isolated stones.
Close to the pole shadowing is particularly important, so that only the tops of spheres are illuminated, creating a negligible TYORP.
This puts limitations on the applicability of the model used in this article, but these limitations are of lesser importance
as polar regions have relatively small area, low illumination levels, and in general smaller lever arms,
all of which minimizes their contribution to the total TYORP effect on an asteroid.

Now consider an asteroid with obliquity $\epsilon=90^\circ$, as in the right-hand panel of Figure \ref{fig:illustration}.
At spring and vernal equinox the boulder A lying at the equator creates the same TYORP as in the previous case with $\epsilon=0^\circ$,
while at solstices it is continuously illuminated from one side (either North or South) and creates no TYORP.
During a year different intermediates between these two regimes occur, 
and the year average should be about one half of the maximal YORP for $\epsilon=0^\circ$, in accordance with Figure \ref{fig:plots}.
Stone B, lying close to the north pole, is shadowed during polar night, in autumn and in winter.
Shortly after spring equinox and shortly before vernal equinox the stone creates the same TYORP as a stone in a polar region of an asteroid with $\epsilon=0^\circ$,
while around summer solstice the stone is perpetually illuminated from above, and creates no TYORP.
The mean of about a half of the maximum TYORP for only half a year means the average TYORP for stone B 
should be about four times smaller than TYORP for a stone in the same position at an asteroid with $\epsilon=0^\circ$.

To obtain the overall torque $T_z$ experienced by an asteroid, one must integrate TYORP over its surface.
It is convenient to introduce the non-dimensional torque $\tau_z$ 
by dividing $T_z$ over $(1-A)\Phi_\odot r^3/c$,
where $c$ is the speed of light, and $r$ is the equivalent radius of the asteroid \citep{golubov12}.
The overall torque also depends on the number of boulders on the surface.
Neglecting mutual shadowing of the boulders, $\tau_z$ is proportional to their density on the surface of the asteroid $n$,
or to the fraction of the surface area occupied by the boulders, $f=\pi R^2 n$.
We compute $\tau_z/f$ for ellipsoidal asteroids as described in Appendix \ref{sec:integration},
and plot the results in Figure \ref{fig:total}.
We find that the TYORP effect for three asteroids with different shapes is almost the same.
The approximate independence of $\tau_z$ was predicted using a simplified analytical model in Appendix B of \cite{golubov14}.
For $\epsilon=0^\circ$ the overall TYORP is nearly twice as big as for $\epsilon=90^\circ$.
This is in a good agreement with the lower two panels in Figure \ref{fig:plots},
where the same roughly two-fold difference between $\epsilon=0^\circ$ and $\epsilon=90^\circ$ is present at low latitudes,
and it is the low latitudes which contribute the most to $\tau_z$ due to their larger area, larger lever arm, and larger $\mdp$.
A simple analytic dependence on latitude with $\tau_z$ proportional to $1+\cos^2 \epsilon$ gives a good fit for a spherical asteroid, 
and also a reasonable estimate for ellipsoidal asteroids (grey line in Figure \ref{fig:total}).
Finally, note that $\tau_z/f$ is nearly 9 times larger than $\mdp$ at the equator, and also agrees with Appendix B of \cite{golubov14}, 
which predicted the factor between $\tau_z/f$ and $\mdp$ to be about 9.

\section{Conclusions}
\begin{figure}
% GNUPLOT: LaTeX picture with Postscript
\begingroup
  \makeatletter
  \providecommand\color[2][]{%
    \GenericError{(gnuplot) \space\space\space\@spaces}{%
      Package color not loaded in conjunction with
      terminal option `colourtext'%
    }{See the gnuplot documentation for explanation.%
    }{Either use 'blacktext' in gnuplot or load the package
      color.sty in LaTeX.}%
    \renewcommand\color[2][]{}%
  }%
  \providecommand\includegraphics[2][]{%
    \GenericError{(gnuplot) \space\space\space\@spaces}{%
      Package graphicx or graphics not loaded%
    }{See the gnuplot documentation for explanation.%
    }{The gnuplot epslatex terminal needs graphicx.sty or graphics.sty.}%
    \renewcommand\includegraphics[2][]{}%
  }%
  \providecommand\rotatebox[2]{#2}%
  \@ifundefined{ifGPcolor}{%
    \newif\ifGPcolor
    \GPcolortrue
  }{}%
  \@ifundefined{ifGPblacktext}{%
    \newif\ifGPblacktext
    \GPblacktextfalse
  }{}%
  % define a \g@addto@macro without @ in the name:
  \let\gplgaddtomacro\g@addto@macro
  % define empty templates for all commands taking text:
  \gdef\gplbacktext{}%
  \gdef\gplfronttext{}%
  \makeatother
  \ifGPblacktext
    % no textcolor at all
    \def\colorrgb#1{}%
    \def\colorgray#1{}%
  \else
    % gray or color?
    \ifGPcolor
      \def\colorrgb#1{\color[rgb]{#1}}%
      \def\colorgray#1{\color[gray]{#1}}%
      \expandafter\def\csname LTw\endcsname{\color{white}}%
      \expandafter\def\csname LTb\endcsname{\color{black}}%
      \expandafter\def\csname LTa\endcsname{\color{black}}%
      \expandafter\def\csname LT0\endcsname{\color[rgb]{1,0,0}}%
      \expandafter\def\csname LT1\endcsname{\color[rgb]{0,1,0}}%
      \expandafter\def\csname LT2\endcsname{\color[rgb]{0,0,1}}%
      \expandafter\def\csname LT3\endcsname{\color[rgb]{1,0,1}}%
      \expandafter\def\csname LT4\endcsname{\color[rgb]{0,1,1}}%
      \expandafter\def\csname LT5\endcsname{\color[rgb]{1,1,0}}%
      \expandafter\def\csname LT6\endcsname{\color[rgb]{0,0,0}}%
      \expandafter\def\csname LT7\endcsname{\color[rgb]{1,0.3,0}}%
      \expandafter\def\csname LT8\endcsname{\color[rgb]{0.5,0.5,0.5}}%
    \else
      % gray
      \def\colorrgb#1{\color{black}}%
      \def\colorgray#1{\color[gray]{#1}}%
      \expandafter\def\csname LTw\endcsname{\color{white}}%
      \expandafter\def\csname LTb\endcsname{\color{black}}%
      \expandafter\def\csname LTa\endcsname{\color{black}}%
      \expandafter\def\csname LT0\endcsname{\color{black}}%
      \expandafter\def\csname LT1\endcsname{\color{black}}%
      \expandafter\def\csname LT2\endcsname{\color{black}}%
      \expandafter\def\csname LT3\endcsname{\color{black}}%
      \expandafter\def\csname LT4\endcsname{\color{black}}%
      \expandafter\def\csname LT5\endcsname{\color{black}}%
      \expandafter\def\csname LT6\endcsname{\color{black}}%
      \expandafter\def\csname LT7\endcsname{\color{black}}%
      \expandafter\def\csname LT8\endcsname{\color{black}}%
    \fi
  \fi
    \setlength{\unitlength}{0.0500bp}%
    \ifx\gptboxheight\undefined%
      \newlength{\gptboxheight}%
      \newlength{\gptboxwidth}%
      \newsavebox{\gptboxtext}%
    \fi%
    \setlength{\fboxrule}{0.5pt}%
    \setlength{\fboxsep}{1pt}%
\begin{picture}(5040.00,3600.00)%
    \gplgaddtomacro\gplbacktext{%
      \csname LTb\endcsname%
      \put(858,704){\makebox(0,0)[r]{\strut{}$0$}}%
      \put(858,1174){\makebox(0,0)[r]{\strut{}$0.005$}}%
      \put(858,1644){\makebox(0,0)[r]{\strut{}$0.01$}}%
      \put(858,2113){\makebox(0,0)[r]{\strut{}$0.015$}}%
      \put(858,2583){\makebox(0,0)[r]{\strut{}$0.02$}}%
      \put(858,3053){\makebox(0,0)[r]{\strut{}$0.025$}}%
      \put(990,484){\makebox(0,0){\strut{}$0$}}%
      \put(1396,484){\makebox(0,0){\strut{}$10$}}%
      \put(1802,484){\makebox(0,0){\strut{}$20$}}%
      \put(2208,484){\makebox(0,0){\strut{}$30$}}%
      \put(2614,484){\makebox(0,0){\strut{}$40$}}%
      \put(3019,484){\makebox(0,0){\strut{}$50$}}%
      \put(3425,484){\makebox(0,0){\strut{}$60$}}%
      \put(3831,484){\makebox(0,0){\strut{}$70$}}%
      \put(4237,484){\makebox(0,0){\strut{}$80$}}%
      \put(4643,484){\makebox(0,0){\strut{}$90$}}%
    }%
    \gplgaddtomacro\gplfronttext{%
      \csname LTb\endcsname%
      \put(220,2019){\rotatebox{-270}{\makebox(0,0){\strut{}$\tau_z/f$}}}%
      \put(2816,154){\makebox(0,0){\strut{}$\epsilon$ [$^\circ$]}}%
      \csname LTb\endcsname%
      \put(3656,3162){\makebox(0,0)[r]{\strut{}1:1:1}}%
      \csname LTb\endcsname%
      \put(3656,2942){\makebox(0,0)[r]{\strut{}2:1:1}}%
      \csname LTb\endcsname%
      \put(3656,2722){\makebox(0,0)[r]{\strut{}2:2:1}}%
    }%
    \gplbacktext
    \put(0,0){\includegraphics{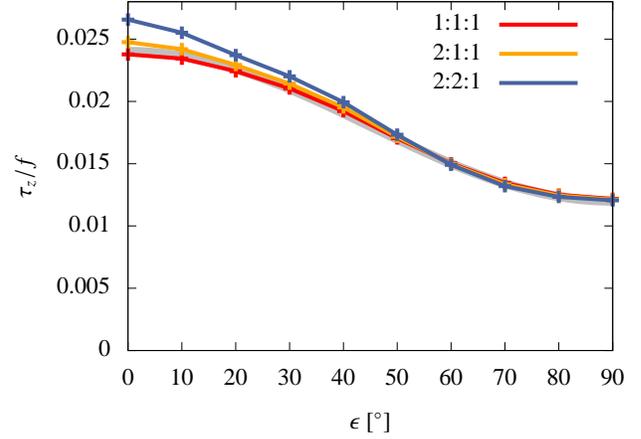}}%
    \gplfronttext
  \end{picture}%
\endgroup\caption{Non-dimensional surface-integrated TYORP torque as a function of obliquity. 
Different colors correspond to different axis ratios of ellipsoidal asteroids.
Grey line corresponds to the formula $0.012(1+\cos^2 \epsilon)$, which well approximates TYORP for a sphere.}
\label{fig:total}
\end{figure}
We find the TYORP effect to operate for non-zero obliquities, making it a truly general effect.
At high obliquities TYORP only gets a factor of a few smaller.
In all the cases tested, TYORP remains positive (see Figure \ref{fig:plots}),
and even in all seasons the day-averaged contribution to TYORP is also positive in all our simulations (see Figure \ref{fig:orbital_variations}).
The numerical value of TYORP for different latitudes and obliquities has a clear physical explanation 
(see Figure \ref{fig:illustration} and its discussion in the text).
The size of boulders and thermal parameters look to be separable from the latitude and obliquity,
so that their alteration only causes a rescaling of the plots as a whole.
The surface-integrated TYORP torque for an ellipsoidal asteroid at high obliquities is about two times smaller than for low obliquities,
with the law $\tau_z\propto 1+\cos^2 \epsilon$ providing a crude estimate (Figure \ref{fig:total}).
Its value and its weak dependence on the shape of the asteroid are in good agreement with the rough analytic estimates by \cite{golubov14}.
The development of these simple models for the TYORP effect enable it to be generally incorporated into more detailed evolutionary models of asteroid rotation when subject to solar illumination effects. 

\begin{acknowledgements}
O.G. and D.J.S. acknowledge support from NASA grant NNX14AL16G from the Near-Earth Object Observation Program and from grants from NASA’s SSERVI Institute.
\end{acknowledgements}

\appendix
\section{Sun coordinates}
\label{sec:coors}
Figure\,\ref{fig:coors} shows different coordinate systems.
Sun in a reference frame co-moving with the asteroid has coordinates:
\begin{align}
    x &= \cos \upsilon \,,\\ 
    y &= \cos \epsilon \sin \upsilon \,,\\
    z &= \sin \epsilon \sin \upsilon\,,
    \label{eq: }
\end{align}
where $\epsilon$ is the obliquity and $\upsilon$ is the angle between the point of the equinox and the direction towards the Sun.
In a co-rotating frame, we have:
%\begin{align}
%     x' &=  x \cos \phi +  y \sin \phi \,,\\
%     y' &= - x \sin \phi + y \cos \phi \,,\\
%     z' &=  z\,,
%    \label{eq: }
%\end{align}
\begin{align}
    x' &=  \cos\upsilon \cos \phi + \cos\epsilon\sin\upsilon\sin \phi \,,\\
    y' &= -\cos\upsilon\sin\phi + \cos\epsilon\sin\upsilon\cos\phi \,, \\
    z' &= \sin\epsilon\sin \upsilon\,,
        \label{eq: }
\end{align}
where $\phi = \omega t$, $\omega$ being the rotational frequency.
Finally, we get the Sun coordinates in a local coordinate system of a boulder
by rotating around the $y'$ axis by the latitude $\psi$:
\begin{align}
    x'' &= (\cos\upsilon \cos \phi + \cos\epsilon\sin\upsilon\sin \phi)\cos\psi + \sin\epsilon\sin \upsilon \sin \psi \\
    y'' &=-\cos\upsilon\sin\phi + \cos\epsilon\sin\upsilon\cos\phi\\
    z'' &= -(\cos\upsilon \cos \phi + \cos\epsilon\sin\upsilon\sin \phi)\sin\psi + 
\sin\epsilon\sin \upsilon\cos \psi
\end{align}
In this local coordinate system, $\vec x''$ axis is a local (outward) normal, 
$\vec z''$ has a direction of a local meridian and $\vec y''$ completes a 
right-handed orthonormal system. Lastly, we simply rename axes so that the $z$ is now the local normal, for convenience.
\begin{figure}
    \imgcoors
    \caption{{Different coordinate systems used in the article and angles between their axes.}}
    \label{fig:coors}
\end{figure}

\section{Integration of TYORP over the surface}
\label{sec:integration}
Let the asteroid be triaxial ellipsoid with axes $a\ge b\ge c$.
We parameterize radius vector belonging to its surface as
\begin{equation}
    \mathbf{r} = (a\sin\theta \cos \alpha, b\sin\theta \sin \alpha, c\cos\theta)\,,
        \label{eq: }
\end{equation}
where the parameters $\theta$ and $\alpha$ are constrained to the ranges $0\le\theta\le\pi$ and $0\le\alpha\le 2\pi$.
Then we take a small surface element, corresponding to small changes of the parameters, $\Delta\theta$ and $\Delta\alpha$.
Two edges of this area are found via partial derivatives of $\mathbf{r}$ over $\theta$ and $\alpha$,
and their vector product gives the vector surface area,
\begin{equation}
    \mathbf{\Delta S} = (bc\sin\theta\cos\alpha,ac\sin\theta\sin\alpha,ab\cos\theta)\sin\theta\,\Delta\theta\,\Delta\alpha\,.
        \label{eq: }
\end{equation}
We find the surface area from the length of this vector, and the latitude $\psi$ from its orientation.
(The latter is necessary to compute $\mdp$.)

To find the lever arm of the TYORP force, we project $\mathbf{r}$ and $\mathbf{\Delta S}$ over the equatorial plane, 
construct a line in this plane crossing through the projection of $\mathbf{r}$ and perpendicular to the projection of $\mathbf{\Delta S}$,
and find the distance from the origin to this line.
Thus we get the lever arm
\begin{equation}
    L = \frac{ab\sin\theta}{\sqrt{a^2\sin^2\alpha+b^2\cos^2\alpha}}\,.
        \label{eq: }
\end{equation}

Finally, the TYORP torque is obtained by numerically adding up the torques created by all surface elements,
\begin{equation}
    T_z = \frac{(1-A)\Phi_\odot f}{C}\sum \mdp\, L\, \Delta S \,,
        \label{eq: }
\end{equation}
where $C$ is the speed of light, and $f$ is the fraction of the surface occupied by the stones.

Values of $\phi$ are different for different facets, and $\mdp$ depends on them.
As we have pre-computed $\mdp$ only for a small set of values of $\phi$, we interpolate between the points using Lagrange polynomials.
It is intentionally constructed to be symmetric with respect to the equatorial plane, 
which makes the obtained dependence more physical and the final result more accurate.

The final TYORP torque is non-dimensionalized as
\begin{equation}
    \tau_z = \frac{T_z C}{(1-A)\Phi_\odot r^3} = \frac{f}{abc}\sum \mdp\, L\, \Delta S \,.
        \label{eq: }
\end{equation}
% WARNING
%-------------------------------------------------------------------
% Please note that we have included the references to the file aa.dem in
% order to compile it, but we ask you to:
%
% - use BibTeX with the regular commands:
%   \bibliographystyle{aa} % style aa.bst
%   \bibliography{Yourfile} % your references Yourfile.bib
%
% - join the .bib files when you upload your source files
%-------------------------------------------------------------------

\end{document}